\makeatletter
\@ifundefined{@parse@version@dash}{%
\def\@parse@version#1{\@parse@version@0#1}
\def\@parse@version@#1/#2/#3#4#5\@nil{%
\@parse@version@dash#1-#2-#3#4\@nil}
\def\@parse@version@dash#1-#2-#3#4#5\@nil{%
  \if\relax#2\relax\else#1\fi#2#3#4 }
}{}
\makeatother
\documentclass[%
 reprint,
superscriptaddress,
 amsmath,amssymb,
 aps,
]{revtex4-2}

\usepackage{graphicx}
\usepackage{dcolumn}
\usepackage{bm}

\usepackage{xcolor}
\usepackage{hyperref}
\hypersetup{colorlinks=true, allcolors={blue},breaklinks=true}

\begin{document}

\preprint{APS/123-QED}

\title{ViPErLEED package II: Spot tracking, extraction and processing of \textit{I}(\textit{V}) curves}

\author{Michael Schmid}
\affiliation{%
 Institute of Applied Physics, TU Wien, Vienna, Austria
}%

\author{Florian Kraushofer}
\affiliation{%
 Institute of Applied Physics, TU Wien, Vienna, Austria
}%
\affiliation{%
Department of Chemistry, TUM School of Natural Sciences, Technical University of Munich, D-85748 Garching bei M\"unchen, Germany
}%

\author{Alexander M. Imre}
\affiliation{%
 Institute of Applied Physics, TU Wien, Vienna, Austria
}%

\author{Tilman Ki{\ss}linger}
\affiliation{%
Solid State Physics, Friedrich-Alexander-Universit{\"a}t Erlangen-N{\"u}rnberg, Erlangen, Germany
}%

\author{Lutz Hammer}
\affiliation{%
Solid State Physics, Friedrich-Alexander-Universit{\"a}t Erlangen-N{\"u}rnberg, Erlangen, Germany
}%

\author{Ulrike Diebold}
\affiliation{%
 Institute of Applied Physics, TU Wien, Vienna, Austria
}%

\author{Michele Riva}
\affiliation{%
 Institute of Applied Physics, TU Wien, Vienna, Austria
}%

\date{\today}

\begin{abstract}
As part of the ViPErLEED project (Vienna package for Erlangen LEED, low-energy electron diffraction), computer programs have been developed for facile and user-friendly data extraction from movies of LEED images. The programs make use of some concepts from astronomical image processing and analysis. As a first step, flat-field and dark-frame corrections reduce the effects of inhomogeneities of the camera and screen. In a second step, for identifying all diffraction maxima (``spots''), it is sufficient to manually mark and label a single spot or very few spots. Then the program can automatically identify all other spots and determine the distortions of the image. This forms the basis for automatic spot tracking (following the ``beams'' as they move across the LEED screen) and intensity measurement. Even for complex structures with hundreds to a few thousand diffraction beams, this step takes less than a  minute.  The package also includes a program for further processing of these $I(V)$ curves (averaging of equivalent beams, manual and/or automatic selection, smoothing) as well as several utilities. The software is  implemented as a set of plugins for the public-domain image processing program {\sc ImageJ} and provided as an open-source package.
\end{abstract}

\maketitle

\section{\label{sec:intro}Introduction}

Analysis of energy-dependent low-energy electron diffraction intensities [LEED $I(V)$ data, also named $I(E)$] is the oldest technique for obtaining high-accuracy data in surface crystallography and has the advantage that it requires rather simple instrumentation (LEED optics), which is available in many ultrahigh-vacuum surface science systems \cite{van_hove_leed_1986,van_hove_ssr_1993,heinz_review_1998,held_bunsen_2010,heinz_electron_2013,fauster_surface_2020}. 
LEED $I(V)$ analysis is based on the comparison of calculated diffraction intensities $I$ as a function of the kinetic energy $E$ of the electrons (or acceleration voltage $V$) with the experimental ones.  The agreement between calculated and experimental $I(V)$ curves is described by an $R$ factor, such as Pendry's $R$ factor $R_\mathrm P$ \cite{pendry_jphysc_1980}, which takes values between 0 for perfect agreement and 1 for uncorrelated curves. (Higher values up to 2 can in principle occur for anti-correlated curves, but are rare.)

In most LEED $I(V)$ studies, symmetry-equivalent beams are averaged
\footnote{Most LEED $I(V)$ studies use normal incidence. Since LEED $I(V)$ spectra depend sensitively on the incidence angle, one can verify normal incidence by comparing the $I(V)$ curves of symmetry-equivalent beams, provided that the surface has sufficient symmetry (at least a rotation axis). Advantages of normal incidence are (i) noise reduction by averaging of symmetry-equivalent beams (also reduction of errors due to residual misalignment) and (ii) the exact incidence angle is known. Off-normal incidence usually requires to handle the exact incidence angle as one or two additional fit parameter(s) in the structure search.}.
The sum of the energy ranges of all the resulting inequivalent beams that enter the analysis is usually named the size of the experimental database. It is well known that care must be taken to provide a sufficiently large experimental database for the structure search. Increasing the size of the experimental data base lowers the risk of the structural analysis becoming stuck in a local minimum of the $R$ factor, improves the accuracy, and increases the trustworthiness of the final result \cite{schmidt_jpcm_2002,kisslinger_prb_2023}. For obtaining a large database, it is desirable to obtain intensity data with sufficient quality not only for bright spots but also for the weak ones. Therefore, a major goal of both the data acquisition and the analysis should be minimizing the noise.

Acquisition and analysis of experimental LEED $I(V)$ data is not only useful for structure determination. $I(V)$ curves are also valuable as fingerprints of structures, especially in cases where different surface structures share the same qualitative appearance of the LEED pattern. This is the case, for instance, if two different terminations of the same bulk crystal have a $(1\times 1)$ LEED pattern, or two different adsorbate coverages lead to the same superstructure. In such a case, LEED $I(V)$ data can verify that a surface preparation can be reproduced. This is useful if other methods to distinguish these two structures are not available in a given vacuum system, or these other methods are not sensitive enough to detect the difference.

The ViPErLEED (Vienna Package for Erlangen LEED) project aims at drastically reducing the effort for LEED $I(V)$ studies, both on the computational and on the experimental side. The package consists of (i) hardware and software for data acquisition \cite{viperleed-acq}, (ii) software for extracting $I(V)$ curves from the experimental data, as well as (iii) software for calculation of $I(V)$ curves for a given structure and structure optimization, by minimizing the difference between the calculated and experimental $I(V)$ data \cite{viperleed-calc}. Part (ii) is the topic of this paper.

Experimentally, $I(V)$ curves are obtained by acquiring images of the LEED screen with a digital camera for a range of energies (usually, several hundred electronvolts, with 0.5 or 1\,eV steps). This results in so-called LEED movies, where the diffraction maxima (the ``spots'') move radially, in the ideal case with a distance from the (0,0) spot proportional to $1/\sqrt{E}$. These LEED movies are processed by following the motion of the diffraction maxima with energy (spot tracking) and evaluation of the intensity of each diffraction maximum (each ``beam'') as a function of energy --- the $I(V)$ curves. For other types of LEED investigations, it is also useful to determine the beam intensities over time or temperature at a fixed energy (e.g., for studying phase transitions); a program for the analysis of LEED intensities should also provide this option.
 
Commercial programs for extraction of $I(V)$ curves from LEED movies usually require selecting each diffraction maximum manually and are often restricted to rectangular regions of interest (ROIs) for intensity integration. Integer ROI coordinates can also lead to jumps of the measured intensity when the ROI moves by a single pixel. Some older programs are also restricted to 8-bit images, and do not take advantage of the high dynamic range of modern cameras (in our experience, about 13--14 bits with the Sony IMX174 sensor and 2$\times$2 binning of pixels).
Among developments by scientific groups, the EasyLEED program by Mayer \emph{et al.}\  \cite{mayer_compphyscomm_2012,url_easyleed} is probably the most suitable development in this field. It is based on a Kalman filter for spot tracking and fitting Gaussians for intensity measurement. This open-source program requires manually selecting each diffraction maximum for measurement, which is a time-consuming and potentially error-prone task in the case of complex superstructures.
The work of Sojka \emph{et al.}\ \cite{sojka_determination_2013,sojka_ultramic_2013, url_leedcal} is based on carefully modeling the relation between the reciprocal lattice and the position on the LEED screen. After a manual step of roughly superimposing the experimentally measured and ideal lattice, this makes it possible to automatically assign $(h,k)$ indices to each spot. While this program is mainly motivated by the desire for accurate measurements of positions in reciprocal space, it could also be extended for $I(V)$ measurements. To our knowledge, though, currently no full solution for $I(V)$ curve extraction based on this program is available.

One problem in obtaining high-quality LEED $I(V)$ data comes from the grids of the LEED optics. There are at least two grids, a grounded grid facing the sample and a suppressor grid at negative voltage that repels electrons that have undergone substantial energy losses by inelastic scattering. It is more common to have three grids, and four grids are used in LEED optics that also serve as retarding-field analyzers \cite{fauster_surface_2020}. The grids absorb diffracted electrons hitting a grid wire, and moir\'e effects can occur from the stacking of differently rotated grids, which results in a spatially inhomogeneous transmission. In addition, further inhomogeneities can result from particles on the grids and dust particles on the camera sensor. In MCP (microchannel plate)-LEED systems, the MCP contributes to the inhomogeneous response. The grids also slightly deflect the electrons, which further complicates the problem. The current work shows how these issues can be mitigated by suitable calibration images (dark screen, flat field).

Our set of programs was written with the aims of (i) making the extraction of $I(V)$ curves as user-friendly as possible, and (ii) obtaining the best data quality with respect to noise and artifacts. The program package is written in Java and based on the public-domain image processing program {\sc ImageJ} \cite{schneider_imagej_2012}, which ensures good performance and operation on all major operating systems (Windows, Linux and MacOS). Details on the installation and use of the programs are provided in the Supplemental Material \cite{supporting}; updates will be published on GitHub \footnote{\url{https://github.com/viperleed/viperleed-imagej}}.

\section{\label{sec:program}Program description}

\subsection{\label{sec:input}Data input}

\begin{figure*}
\includegraphics[width=17cm]{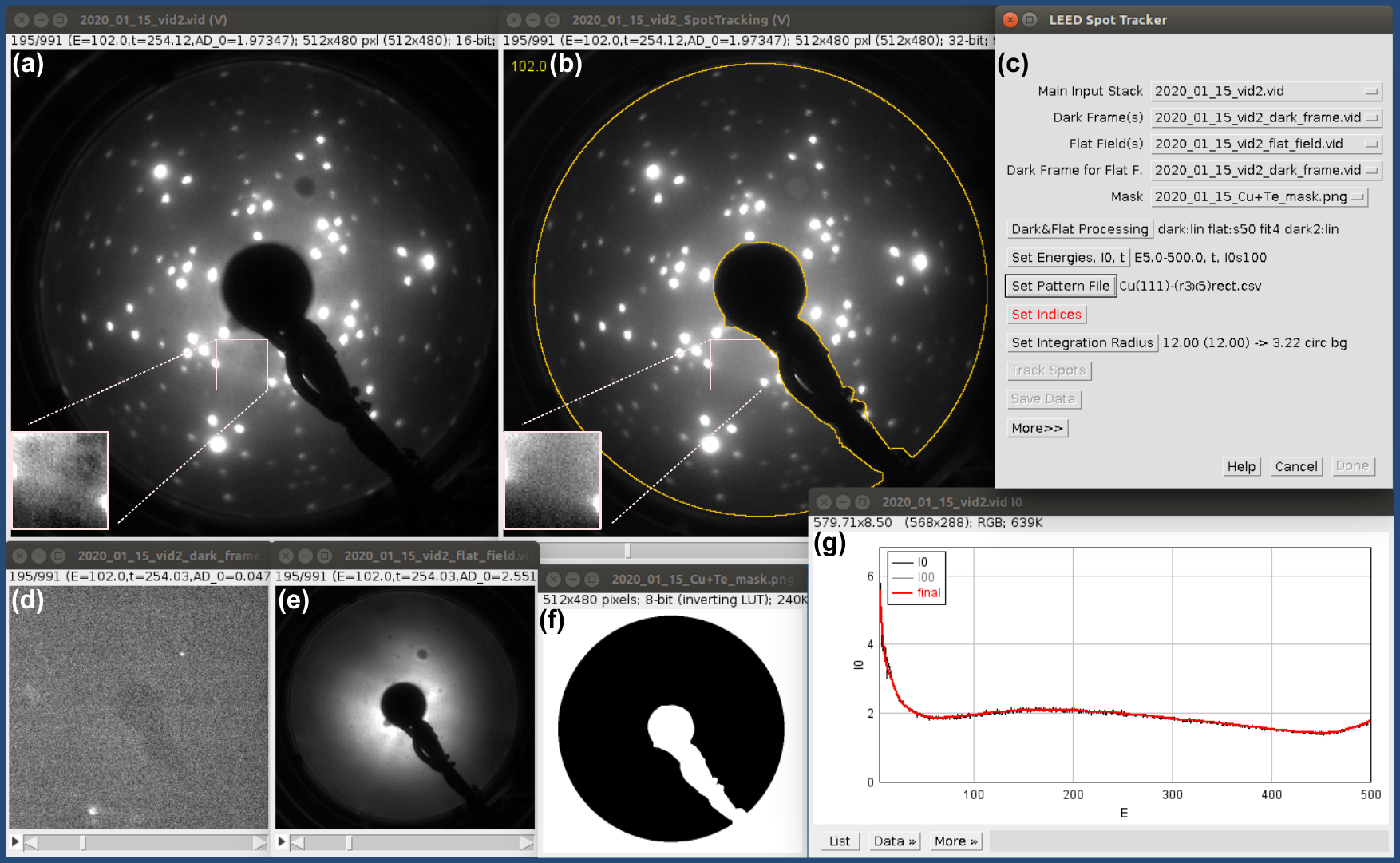}
\caption{\label{fig:spottracker}The graphical user interface of the ViPErLEED spot tracker (here under a Linux operating system), with (a) the input image stack, (b) the processed stack after dark-frame and flat-field correction, and (c) the main spot-tracker panel. The image stacks (d) and (e) are the dark frames and flat field, respectively, (f) is the mask of the usable screen area [also visible as orange outline in (b)], and (g) is a plot of the raw and smoothed beam current $I_0$ (available via the ``Set Energies, I0, t'' button; the $I_{00}$ line is invisible because it coincides with the $x$ axis at this scale). The red item in (c) indicates that user input is required. In (a) and (b), the LEED images of the Cu(111)-$(5\times\sqrt{3}_\text{rect})$-4Te structure \cite{kisslinger_prb_2021} are displayed with high contrast, leading to saturation of the bright spots. The magnified and contrast-enhanced insets in (a) and (b) show the improvement of the background uniformity with the dark-frame and flat-field correction.}
\end{figure*}

\emph{Input files} ---
The main {\sc ImageJ} plugin is the {\em Spot Tracker} (Fig.\ \ref{fig:spottracker}). The main input are LEED movies (named {\em image stacks} in {\sc ImageJ}); these can be opened by appropriate {\sc ImageJ} commands (\texttt{File$>$Open} or \texttt{File$>$Import$>$Image Sequence}), thus any image format that can be read by {\sc ImageJ} or one of its plugins can be used.
The plugin package can also open \texttt{.zip} archives (containing images and an index file with the list of images and metadata) created by the ViPErLEED data acquisition \cite{viperleed-acq}, as well as \texttt{.vid} files of the ``AIDA'' (Automatic Image and Data Acquisition) EE2000/EE2010 program \cite{url_ee2000}. For these formats, the metadata such as energy, time, beam current $I_0$ and additional analog input channels are also read. When reading a collection of single images with  \texttt{File$>$Import$>$Image Sequence}, these data can be decoded from the file names. The ``Set Energies I0, t'' button of the spot tracker panel also includes an option to enter these values as a linear function or read them from an {\sc ImageJ} table. (In {\sc ImageJ}, opening a comma- or tab-delimited file, \texttt{.csv} or \texttt{.tsv}, creates a table).
It is also possible to specify an independent variable other than the energy. This is useful for intensity measurements during a phase transition as a function of time or temperature, at fixed energy.
%

\emph{The mask} ---
The spot tracker requires that the user provides a {\em mask}, which is an image that defines the usable area of the LEED screen [Fig.\ \ref{fig:spottracker}(f)]. This is a binary image, implemented in {\sc ImageJ} as an 8-bit image with only two different pixel values occurring: 255 (black) for the foreground (usable) area and 0 (white) for the unused area. A utility for creation of such an image is available via the ``More$\gg$'' button of the spot-tracker panel. The standard {\sc ImageJ} selection and image-modification commands can be used to edit the mask.

\subsection{\label{sec:preprocessing}Dark-frame and flat-field correction}

The spot tracker has provisions for dark-frame and flat-field correction of the LEED images, which can substantially improve the data quality. These corrections are standard in astronomical image processing \cite{buil_ccd_1991} and in some applications of light microscopy, but not widely used in the LEED community. The aim of these corrections is reducing the effect of inhomogeneities of the LEED optics (grids, and microchannel plate, if any) and camera as well as subtracting background illumination, for instance, from the filament of the electron source. In the standard method, intensities are corrected pixel by pixel,
\begin{equation}
  I_\mathrm{corr} = \frac{I_\mathrm{main} - I_\mathrm{dark}}{I_\mathrm{flat} - I_\mathrm{dark}}\ ,\label{eq:darkflat}
\end{equation}
where $I_\mathrm{main}$ is the pixel intensity of the LEED image with the diffraction pattern and $I_\mathrm{dark}$ is the pixel intensity of a dark frame. The dark frame is an image obtained without electrons reflected at the sample. $I_\mathrm{flat}$ is the pixel intensity of the flat field, which is an image with uniform illumination.

The \emph{dark frame} is best obtained with the same filament current and screen voltage as the main image, but with a highly negative Wehnelt voltage to suppress all electrons. All other settings (exposure time, camera gain) should be the same as for the main LEED images. This ensures that the intensity of any stray light of the filament is the same for both the main LEED images and the dark frames, and, hence, this background intensity will be subtracted (together with the dark current of the camera). In some cases, there can be also a background due to field-emitted electrons from asperities on a grid [such as the bright spots in Fig.\ \ref{fig:spottracker}(d)], which will be subtracted by this procedure. When recording a full movie of energy-dependent dark frames and the LEED electronics provide a beam current ($I_0$) output, the $I_0$ measurement acquired with this movie conveniently provides the energy-dependent offset $I_{00}$ of the beam current (cf.\ Sec.\  \ref{sec:I0}).

For obtaining a \emph{flat field}, one should have uniform illumination of the LEED screen with electrons coming from the same position as the reflected electrons forming the usual LEED image. This is not easy to achieve. The best option we found is placing a polycrystalline surface (e.g., the sample holder
\footnote{Note that annealing polycrystalline materials can lead to grain growth and, thus, the appearance of LEED spots for such materials. To ensure that this is not the case, it is a good practice to inspect the stack of flat-field images as it will be applied to the main input. This image stack, processed with the appropriate dark frame, the normalization polynomial in Eq.\ (\ref{eq:darkflatfit}), and averaging for noise reduction, is available with the ``Show processed flat field'' option in the ``Dark\&Flat Processing'' dialog.})
acting as a diffuse scatterer at the same position as the sample \cite{koller_surfsci_2002}. The distance from the electron source to the surface must be the same for the main LEED $I(V)$ movie and the flat field. In other words, both, the sample and the polycrystalline surface must be exactly in the same plane for the respective measurements. Since the flat-field intensity is spread out over the whole screen, the flat-field images taken with the same settings as the main $I(V)$ movie of the sample might be rather noisy due to low intensity. In that case, one may use a higher beam current and/or longer exposure times than for the main $I(V)$  movie to ensure a sufficiently high intensity. Obtaining flat fields from a polycrystalline surface has the problem of angle-dependent scattering, typically with a maximum at 180$^\circ$ scattering angle (backscattering); see Fig.\ \ref{fig:spottracker}(e) for an example. Thus, the correction in equation (\ref{eq:darkflat}) would introduce a bias, attenuating diffraction intensities near the center compared with those at the periphery.  For standard LEED $I(V)$ experiments, this will also lead to an apparent decrease of intensity towards high energies, because the beams move inwards. We therefore use a modified correction
\begin{equation}
  I_\mathrm{corr} = \left.(I_\mathrm{main} - I_\mathrm{dark}) \middle/ \left(\frac{I_\mathrm{flat} - I_\mathrm{dark2}}{\exp(\sum{a_{ij}x^i y^j})}\right)\right.\ ,\label{eq:darkflatfit}
\end{equation}
where the polynomial $\sum{a_{ij}x^i y^j}$ is a fit to the logarithm of the background-corrected $I_\mathrm{flat} - I_\mathrm{dark2}$ values inside the area defined by the mask. A second-order polynomial would correspond to a 2D Gaussian distribution of the flat-field intensity; typically, we use a 4th-order polynomial for better uniformity of the flat-field correction while still maintaining the high spatial frequencies of the inhomogeneities. As the fit is done in the logarithmic domain we use fit weights proportional to the $I_\mathrm{flat} - I_\mathrm{dark2}$ value; otherwise low values would gain too much weight (especially if the logarithm is highly negative). The flat-field images should have sufficient brightness to ensure low noise; therefore, as mentioned above, the camera settings (gain, exposure time) for recording the flat field might be different from those used for the main LEED movie. In such a case, it will be necessary to obtain a separate set of dark frames with these settings, different from the dark frames for the main $I(V)$ movie. This is indicated by the ``2'' in $I_\mathrm{dark2}$. If the same camera settings are used for the main $I(V)$ movie and for the flat field, the same dark frame(s) can be selected for both (i.e., $I_\mathrm{dark2} = I_\mathrm{dark}$).

The dark frames depend at most weakly on the beam energy. (A weak dependence is possible if field emission from the last grid to the screen causes a background intensity and the voltage between the grid and screen varies with the beam energy.) If the dark frames are energy-independent, it is enough to average over a few dark-frame images to reduce the noise; otherwise a linear fit of each pixel intensity over energy is usually sufficient. These options are accessible via the ``Dark\&Flat Processing'' button of the spot tracker.

In the flat-field images, as mentioned above, the intensity is spread out over the whole screen, which leads to an intensity below that of the spots in the main LEED $I(V)$ movie, and, hence, higher noise. Therefore, noise reduction should be applied by smoothing the pixel intensity vs.\ energy; also this function is available in the ``Dark\&Flat Processing'' options. It requires that the flat fields are acquired as an image stack with the same energy steps as the main LEED image stack. When using energy-dependent flat fields, but not a 2D fit for the flat-field intensity, the ($I_\mathrm{dark2}$-corrected) flat field should be normalized, to avoid influencing the $I(V)$ data by the energy dependence of the diffuse backscattering, which creates the dark-field images.

In our experience, the dark-frame/flat-field correction has a profound impact on the data quality. This is especially true for LEED measurements with the sample at room temperature, where the background from scattering by phonons is high and therefore its variations due to the grid wires and other inhomogeneities of the grids are clearly visible. The improvement of the background uniformity is also evident in LEED movies recorded at low temperature, where the background is low [compare the insets in Figs. \ref{fig:spottracker}(a) and \ref{fig:spottracker}(b)]. The correction especially improves the quality of the $I(V)$ curves of weak spots, where the background fluctuations have a comparably strong impact on the intensity measurements. Unfortunately, the correction cannot fully eliminate the influence of the grids on the beam intensities: The electrons get deflected by the lateral electric-field components of the suppressor grid: the grid meshes act similarly to electrostatic lenses. Since the flat-field correction is based on the position of the diffracted beam on the screen (recorded by the camera), which can deviate from the original direction of the diffracted electrons before they reach the grids, the flat-field intensity distribution at the screen cannot accurately describe intensity variations depending on where the electrons reach the grid. This is mainly a problem with highly focused beams (very sharp spots). The intensity noise caused by the modulation by the grids can be reduced by slightly defocusing the electron beam. This is only possible if the spots are sufficiently far apart for accurate determination of the background (see Sec.\  \ref{sec:spotanalysis}), and the background of inelastically scattered electrons is low. (Otherwise, weak, smeared-out spots will not stand out high enough over the background.) A further method to reduce the noise due to the grid structure is averaging LEED $I(V)$ curves obtained from movies with slightly different azimuthal rotation of the sample (if the sample manipulator allows this) or slightly different distance to the sample (1--2\,mm shift is sufficient; in this case also a flat field should be recorded for each distance).

The flat-field correction also increases the usable screen area near the electron source. Since camera lenses with a large aperture are required for good photon collection efficiency, the outline of the electron source appears blurred in the images because it is out of focus [Fig.\ \ref{fig:spottracker}(a)]. The flat-field correction compensates for the reduced intensity recorded where the screen is partly hidden by the electron source. Thus the mask of usable screen area (see below) can extend closer to the edge of the electron source than without flat-field correction, and the usable energy range of spots disappearing behind the electron source is extended.

\emph{Implementation notes} ---
All operations on the input image stacks are implemented as {\sc ImageJ} \texttt{VirtualStack}s, which means that the processed images are not necessarily kept in memory but rather read from disk and calculated on the fly as required. Only the final result $I_\mathrm{corr}$ is cached in memory as long as there is enough RAM (using the Java \texttt{SoftReference} mechanism and prefetching). This ensures that even very large image stacks can be handled while good performance is achieved when there is sufficient memory.

\subsection{\label{sec:maxima}Distortion correction for identification of the spots}

\emph{Polynomial fits} ---
A LEED pattern is essentially a 2D map of the reciprocal space, with some distortions that come from various sources: The point where the incident beam hits the sample may not exactly coincide with the center of curvature of the LEED grids and screen, the camera is not at infinite distance and not necessarily aligned with the axis of the incident electron beam, the grids and/or screen may deviate from the ideal spherical-cap shape, there may be residual electric and magnetic fields, and the sample may be tilted. Some of these sources of distortions should be clearly minimized when acquiring LEED $I(V)$ data. (Usually normal incidence of the electrons on the sample is desired, and stray fields must be avoided.) Nevertheless, it is not possible to avoid all sources of distortion. Many sources of distortion can be modelled \cite{sojka_determination_2013,sojka_ultramic_2013}, but this is rather cumbersome for the general case. Therefore, we took a more simplistic approach: We fit the pixel coordinates $x, y$ with a polynomial function of the reciprocal-space coordinates $k_x, k_y$,
\begin{equation}
  x = \sum_{i,j; i+j\le N}{a_{ij} k_x^i k_y^j}\ ,\quad
  y = \sum_{i,j; i+j\le N}{b_{ij} k_x^i k_y^j}\ . \label{eq:fitkxky}
\end{equation}
The polynomial order $N$ is chosen adaptively (see below); the maximum order supported is 5\textsuperscript{th} order.  In addition to polynomials with all coefficients up to a given order, the program also includes models where the highest-order terms only depend on the reciprocal-space distance from the (0,0) spot, but other high-order coefficients are left out:
\begin{align}
  x &= \sum_{i,j; i+j\le1}{a_{ij} k_x^i k_y^j} + (k_x^2+k_y^2)(a_{\mathrm{rx}}k_x+a_{\mathrm{ry}}k_y)\nonumber\\
  y &= \sum_{i,j; i+j\le1}{b_{ij} k_x^i k_y^j} + (k_x^2+k_y^2)(b_{\mathrm{rx}}k_x+b_{\mathrm{ry}}k_y)\ , \label{eq:fit13}
\end{align}
and
\begin{align}
  x &= \sum_{i,j; i+j\le3}{a_{ij} k_x^i k_y^j} + (k_x^2+k_y^2)^2(a_{\mathrm{rx}}k_x+a_{\mathrm{ry}}k_y)\nonumber\\
  y &= \sum_{i,j; i+j\le3}{b_{ij} k_x^i k_y^j} + (k_x^2+k_y^2)^2(b_{\mathrm{rx}}k_x+b_{\mathrm{ry}}k_y)\ . \label{eq:fit35}
\end{align}
These two types of fit polynomials are suitable in case of normal incidence and mainly radial distortions. They offer the advantage of handling radial distortions with fewer fit parameters (10 and 24) than the full third- and fifth-order polynomial fits (20 and 42 parameters, respectively), thus they do not require as many spots as the full third- and fifth-order polynomials in Eq.\ \ref{eq:fitkxky}.

\emph{The spot pattern file} ---
Correlating the spots on the screen and the reciprocal-space coordinates requires a list of beams for the structure. This list must be provided as a {\em spot pattern file}. For each beam, it lists the designation and the indices $h$ and $k$, the Cartesian reciprocal-space coordinates $g_x, g_y$ (in arbitrary units) and a beam-group index (symmetry-equivalent beams belong to the same group). This file can be created with the LEED pattern simulator of the ViPErLEED GUI, supplied with the ViPErLEED data acquisition \cite{viperleed-acq}. There, the lattice and overlayer symmetry can be entered manually or taken from the output of the ViPErLEED simulation program \cite{viperleed-calc} (\texttt{experiment-symmetry.ini} file).

\emph{Identification of the spots} ---
In practice, for determination of the fit coefficients of Eq.\ (\ref{eq:fitkxky}), the user has to select an energy where many spots can be seen, preferably including spots near the edges in many different directions from the center. To aid this procedure, spots with sufficient brightness are marked by circles after pressing the ``Set Indices'' button. The spots are found as local maxima with a threshold (for noise suppression), based on the \texttt{Find Maxima} function of {\sc ImageJ}, and their positions are refined as described in Sec.\  \ref{sec:spotanalysis}. In the next step, the user has to select one of these spots and enter its $(h,k)$ indices. If only one spot is known, the program will assume that the (0,0) spot is in the center of the screen (given by the bounding box of the mask area described above) and search for additional spots, starting with those closest in reciprocal space to the initial spots. At first, a purely linear relationship will be tried. Whenever a new spot has been identified, the fit in Eq.\ (\ref{eq:fitkxky}) is repeated with the new spot included. If there are enough spots for obtaining higher-order polynomial coefficients, the program attempts fitting with a higher-order polynomial [including the functions in eqs. (\ref{eq:fit13}) and (\ref{eq:fit35}); in the sequence of increasing number of fit parameters] and uses the higher order if the goodness of fit improves (taking into account that a larger number of fit parameters will reduce the residuals).

Since the polynomials in Eq.\ (\ref{eq:fitkxky}) are not necessarily monotonic, it can happen that the polynomials map high-order spots far outside the screen (or even non-existent at a given energy) to a position inside the LEED screen. If such a position happens to coincide with a lower-order spot (or a defect of the screen that is mistaken as a spot), this will cause misidentification of that spot (or defect). The program therefore contains provisions to discard the calculated positions in such a case: A spot position obtained from Eq.\ (\ref{eq:fitkxky}) is accepted only if the nonlinear terms in the polynomial do not ``deflect'' the direction of spot motion with increasing energy by more than 30$^\circ$ (with respect to the direction calculated from the linear terms). This prevents misidentification of spots in cases where the calculated position folds back to the screen area (like the ``down`` branches of an $x - x^3$ polynomial). For 5\textsuperscript{th}-order polynomials, this condition is not sufficient. For spots actually visible on the screen, the 5th-order terms are only a small correction (even for off-normal incidence). The 5\textsuperscript{th}-order terms  can become large for spots that are actually far outside the screen area, and make the polynomial fold back and forth across the LEED screen (like $x - x^3+x^5/5$, which has a three zeros with positive slope). Therefore, for 5\textsuperscript{th}-order polynomials, it is also required that the derivatives of the respective 4\textsuperscript{th}-order polynomial (without the 5\textsuperscript{th}-order terms) fulfill the same condition as the full polynomial.

For (almost) normal incidence, it is usually sufficient to manually enter the $(h,k)$ indices of one spot; the program then automatically identifies all the others. In some cases, especially far from normal incidence, it may be required to select several spots and enter their $(h,k)$ indices manually. We have successfully tested the program with up to 20$^\circ$ off-normal incidence, where correct identification of all spots usually requires manual input of the $(h,k)$ indices for 3--4 spots. Apart from the spot labels shown on the image, a correct identification can also be inferred from low values of the root-mean-square (rms) residuals of the pixel positions with respect to the polynomial fit, as calculated and displayed by the program. Typical rms residuals are $\lesssim 0.2$\% of the image width (1 pixel for a 512$\times$512 pixel image).

\subsection{\label{sec:spotanalysis}Analysis and intensity evaluation of a single diffraction maximum}

Analysis of a single spot has two major aims, determination of (i) the position and (ii) the intensity. The problem of spot analysis is comparable to photometry of single stars in astronomy, and there are two basic approaches \cite{mighell_photometry_1999}: Aperture photometry and fitting of a point spread function (PSF). The PSF is the intensity distribution that one would obtain for an idealized ($\delta$-like) maximum. In principle, PSF fitting has the potential of better accuracy in terms of both position and intensity, but it requires knowledge of the PSF. (In astronomy, stars are almost perfect $\delta$ functions, all smeared out the same way by atmospheric turbulence and the optics; thus the PSF can be obtained by averaging the images of a few bright stars without nearby background objects.) For LEED diffraction maxima, the PSF cannot be determined for several reasons. (i) Due to deflection of electrons by the suppressor grid and electron capture by grid wires, the spot intensity profiles show modulations caused by the grid. These modulations depend on the position on the screen. (ii) Due to the curvature of the screen, spots are distorted towards the edges of the screen. (iii) On samples with a high step density (step--step separation less than or comparable to the transfer width of the instrument), the width of the spots depends on their index and the energy in a non-trivial way. (Assuming kinetic theory, broadening occurs at out-of-phase conditions \cite{henzler_quantitative_1978}.) (iv) The background due to scattering by phonons is not constant but increases towards the spots \cite{mckinney_thermal_1967}. This makes it difficult to separate the contributions of the spot and the phonon background. Instead of using a pre-determined PSF, one can also use independent 2D Gaussian fit functions for each spot \cite{mayer_compphyscomm_2012}, but this approach becomes difficult at low intensities, where the fit is ill-defined and further complicated in case of a sloping background.

\begin{figure}
\includegraphics[width=8cm]{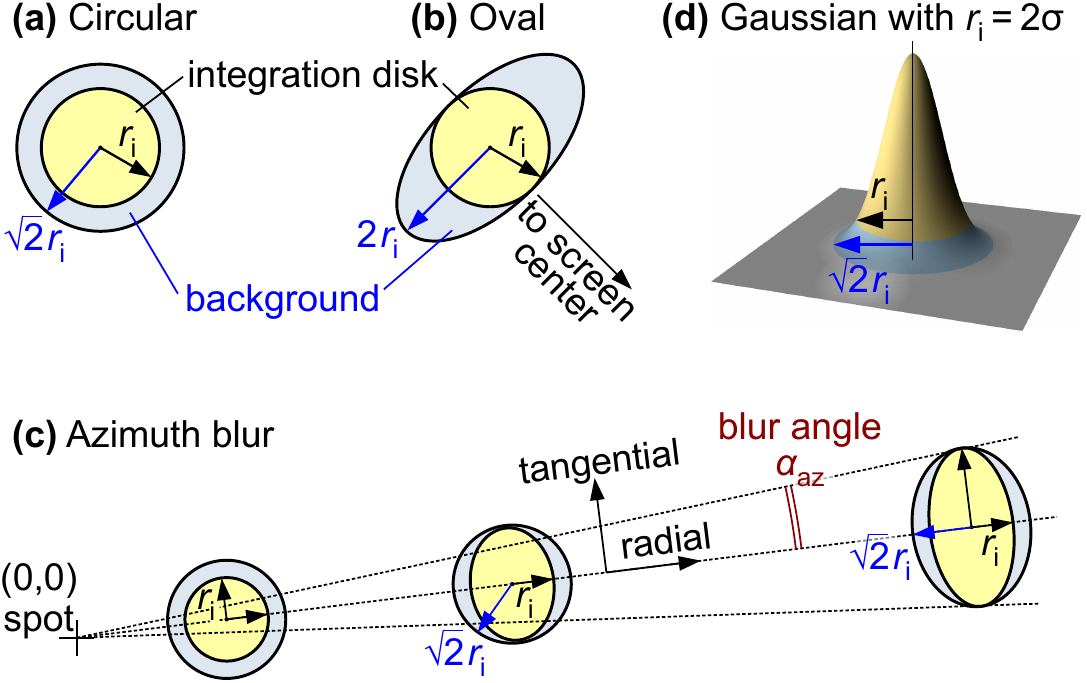}
\caption{\label{fig:aperturePhotometry}Aperture photometry.  Integration disk and area for background intensity evaluation with (a) circular and (b) oval outline of the background area. (c) Shapes of the integration and background areas in mode ``azimuth blur'' for spots with different distances from the (0,0) spot. (d) Illustration of the minimum integration area for LEED spots proposed, assuming a Gaussian profile and annular background (light blue), as shown in panel (a). }
\end{figure}

The other approach to spot analysis is known as aperture photometry (Fig.\ \ref{fig:aperturePhotometry}) \cite{howell_aperture_1989}, and in astronomical image processing it was shown that its accuracy for intensity measurements can be comparable to or even surpass PSF fitting \cite{sonnett_photometry_2013}.  Aperture photometry integrates the intensity over a (usually circular) disk; the background intensity is taken from an annular area around the integration disk. In the most simple case, the average of the pixel intensities in the background area is taken, but other schemes like median, histogram centroid or statistical mode are also common \cite{howell_aperture_1989}.  For the background of LEED spots, different methods of evaluating the background intensity were compared in Ref.\ \onlinecite{roucka_vac_2002}; good results were achieved with fitting a linear or 2\textsuperscript{nd}-order polynomial in $x$ and $y$. The profile of a LEED spot decays rather slowly at large distances $r$ from its center ($1/r$ or $1/r^2$) \cite{mckinney_thermal_1967}. A 2\textsuperscript{nd}-order polynomial does not provide a good description of this decay. Compared to a linear fit, 2\textsuperscript{nd} order also has the disadvantage of more free parameters, which tends to increase the noise. Therefore, the program uses a linear function in $x$ and $y$ to fit the intensity in the background area. This ensures that the measurement is independent of the gradient of the background
\footnote{Instead of fitting a linear background one could simply use the average over the background area if the spot is exactly centered (vanishing first moments over the integration disk after subtraction of the linear background) and the shapes of the integration areas for the spot and the background have at least twofold rotation symmetry around the center. Subtracting the linear background is required for obtaining the spot position via the first moments, and also for the intensity measurement if the spot is not perfectly centered spot or there is an asymmetry of the integration areas. Slight asymmetry can occur due to the spatial quantization (image pixels). We use a one-pixel-wide transition zone where the weight of the background evaluation decreases to zero; this transition zone is not required to be fully inside the foreground area of the mask. If pixels in the transition zone have to be excluded because they are outside the mask foreground area, this also causes asymmetry.}.

In astronomy, where the distance between the stars is often much larger than the size of the PSF, it is common to use an inner radius of the background annulus that is larger than the outer radius of the integration disk for the star intensity. Thus, there is a dead zone in between. For the analysis of LEED spots, we use a few modifications of this scheme. The program offers three different geometries for the integration and background areas.

The first is an \emph{annular background} with the inner radius equal to the radius $r_\mathrm i$ of the integration disk, and an outer radius of $\sqrt{2}r_\mathrm i$ [Fig.\ \ref{fig:aperturePhotometry}(a), named ``circular'' in the program]. Thus, the background area is equal to the integration area. This choice was motivated by the following consideration about noise: If the spot intensity vanishes, the noise obtained in the integration disk and the background annulus (by averaging) are equal. Thus, the spot-minus-background noise is higher by a factor of $\sqrt{2}$ than the noise obtained from integration over the inner disk
\footnote{Since the noise of different pixels is usually uncorrelated, one can use the rules of error propagation to estimate the noise. If the areas of the integration disk and background annulus are the same, at vanishing spot intensity (i.e., without intensity-dependent shot noise), the noise-related errors of the two integrals over these areas will be the same. Thus, the error of the difference of these two integrals equals $\sqrt{2}$ times the error of one of the integrals.}.
If the spot intensity is higher, the influence of the background noise on the $I(V)$ curves will be less, since both the shot noise and spot intensity modulations due to the grid increase with intensity. 
In contrast to astronomical aperture photometry, we do not use a dead zone between the integration disk and the background area. The main reason is the non-uniform background in LEED (see Fig.\ \ref{fig:spottracker}). If the background is a nonlinear function of $x$ and $y$, the non-uniformity induces a background error that increases with increasing radius of the background annulus. The other reason for having a small background annulus is trivial: For complex LEED patterns and high energies, the distance between the spots becomes small, and the background area of one spot must not overlap with the neighboring spots.

The \emph{oval background} is second type of background area available in our program. It does not use a circular outer boundary but rather an ellipse with the semiminor axis equal to the radius of the integration disk and the semimajor axis twice as large [Fig.\ \ref{fig:aperturePhotometry}(b)]. As in the case of the annular background, the inner boundary is given by the integration disk, and the background is averaged over as many pixels as the integration disk of the spot. The major axis of the ellipse is in the tangential direction (we simply take the center of the bounding box of the mask as the center). The main advantage of this background type (named ``oval'' for short) is better suppression of radial variations of the background intensity compared with the annular background. This is especially valuable for some channel-plate LEED optics, where concentric ringlike artifacts in the background intensity occur. The oval background is also valuable for standard LEED instruments, due to the (nonlinear) decrease of background intensity from the center of the screen. An additional bonus of the oval background is a slightly increased usable energy range in most cases: If a spot is close to the outer edge of the LEED screen or the electron source, but its integration disk is still inside the usable  screen area (the mask), an annular background area may reach beyond the screen and prevent a well-defined measurement. The oval background area protrudes only in the tangential direction and can be fully evaluated until the integration disk of the spot touches the edge
\footnote{For spots moving along the arm holding the electron source, which is essentially a radial ``spoke'' (to the bottom right in Fig.\ \ref{fig:spottracker}), the oval background touches that arm before an annular background area with equal area would touch it. This can reduce the amount of data available with the oval background as compared with a circular background.  This case occurs less often than that of spots close to the inner or outer boundary. If it occurs, it often affects only one of several symmetry-equivalent beams, so it does not affect the size of the experimental database of the symmetry-averaged $I(V)$ curves.}.
The oval background is inferior to the annular one in case of very crowded LEED patterns, because neighboring spots will typically enter the oval background area due to its larger extension in the tangential direction before they would affect the annular background. The oval background is also less suitable than the annular one if non-radial background variations are dominating, such as background variations due to short-range order or phonons: The oval background reaches out further than the annular background area, thus it is more sensitive to non-radial background variations that are a nonlinear function of $x$ or $y$.

\emph{Azimuth blur mode} ---
Finally, there are situations where the spots are blurred in the azimuthal direction. This happens in the case of overlayers with poorly determined azimuthal orientation. For this case, we offer an option that is close to the circular geometry in the vicinity of the (0,0) spot, but the integration area becomes elongated in the tangential direction at larger distances from (0,0), see Fig.\ \ref{fig:aperturePhotometry}(c). For simplicity, we use an elliptical integration area, not an arc; this limits the blur angle to small values (a few degrees). In this geometry, one should not measure the background intensity all around the integration ellipse; in the tangential direction the background evaluation area would be too far from the spot center and therefore the measurement would become very sensitive to spatial variations of the background. For high eccentricity, we therefore use the geometry shown at the right side of Fig.\ \ref{fig:aperturePhotometry}(c), where the outer border of the background area is an ellipse touching the integration ellipse at the vertices (in the tangential direction). For this geometry, we take a smaller ratio between the background and integration areas than in the circular and oval case, to limit the influence of nonlinear variations of the background in the radial direction: As soon as the ratio between the major and minor axes of the integration ellipse exceeds $\sqrt{2}$, the minor axis of the outer background border is limited to $\sqrt{2}$ times the minor axis $2r_\mathrm{i}$ of the integration disk. This results in a ratio between background area and integration area of 0.41. For more circular integration ellipses [closer to the (0,0) spot], we use a circular outline of the background, with a radius of $\sqrt{2}$ times the semiminor axis of the integration ellipse, see the two left cases in Fig.\ \ref{fig:aperturePhotometry}(c). This results in our usual ``circular'' geometry close to the (0,0) spot, where azimuthal blurring is negligible.

Both the position of the integration and background areas and their borders are calculated with subpixel accuracy. For integration, pixels in a one-pixel-wide zone at the border are weighted between 1 (inside) and 0 (outside that zone). This avoids jumps that could otherwise occur in case of very sharp spots and small integration areas.

Spot analysis is not only used for intensity measurements but also for determining the exact spot position; this is required for fitting the polynomial model in Sec.\  \ref{sec:maxima} and when tracking the spots (section \ref{sec:tracking}). For this purpose, it is important to fit a linear background in the (oval or annular) background area. After subtraction of this background, we determine the position of the center of mass inside the integration disk. This process is repeated iteratively until convergence (iteration step less than 0.3 pixels) or aborted if the new position deviates too much from the initial guess (this can happen when searching for a spot with vanishing intensity).

In addition to integrated intensity and position, spot analysis also yields an estimate for the spot size derived from the second moments inside the integration disk. The program gives the spot size as standard deviation $\sigma$ assuming a 2D Gaussian
\footnote{For the calculation of $\sigma$ from the moments we use a heuristic correction to take into account that the spot intensity is not fully inside the integration disk. Since experimental LEED spot profiles are non-Gaussian and typically have slowly decaying tails at low intensity \cite{mckinney_thermal_1967}, the spot size is typically overestimated, especially if the integration radius is large.};
the sizes in the radial and tangential directions are given separately.
Finally, we also extract a measure of significance, which depends on the ration of the spot intensity and the standard deviation of the background (after subtraction of a linear background); this value is used to obtain smoothed spot positions (section \ref{sec:tracking}).

\emph{The integration radius} ---
The choice of the integration radius depends on the spot size and whether spots come close to each other at high energies. It has been suggested to use an adaptive integration area that only encompasses the region where the spot can be clearly discerned from the background, thus shrinking the area with decreasing intensity \cite{toofan_rsi_1994}. This approach has the advantage of reducing the noise for weak spots, but it is problematic because it introduces a bias: For weak spots, only very center will be inside the integration area and the remaining intensity discarded. Thus the intensity of weak spots will be underestimated. Therefore, we use an integration disk with a radius that does not depend on the intensity.
For a 2D Gaussian with a standard deviation of $\sigma$, illustrated in Fig.\ \ref{fig:aperturePhotometry}(d), 86\% of the intensity is contained in a circle with a radius of $2\sigma$. Using an annular background as described above, most of the remaining intensity will spill into the background annulus and increase the background, reducing the integral-minus-background measurement to 75\% of the total intensity. As long as the shape of the intensity distribution stays the same, this factor is constant and has no detrimental effect on the $I(V)$ curves, where absolute intensity is not important. While the optimal radius of the integration disk in astronomical photometry is lower ($\approx 1.6 \sigma$, Ref.\ \onlinecite{mighell_photometry_1999}), we consider $2\sigma$ the minimum integration radius for a good LEED intensity measurement. The reason for the difference lies in the fact that astronomy deals with a roughly constant PSF of stars, while the shapes of the LEED diffraction maxima change, due to deflection and capturing of the electrons by the grid. Furthermore, astronomy uses a dead zone between the integration disk and the background annulus, which is impractical for LEED (see above). The grid-related noise increases with decreasing size of the integration area. In our experience, this increase becomes significant at $r_\mathrm i \lesssim 2\sigma$.  Therefore, if the distance between the spots at high energy allows it, the integration radius should be chosen slightly larger than $2\sigma$. On the other hand, for weak spots, the noise of the measured intensities increases with increasing size of the integration disk (the image noise is integrated over a larger area). The signal-to-noise ratio of the measured intensities will typically have a minimum at a radius close to or somewhat larger than $2\sigma$. In addition, as mentioned above, the impact of nonlinear background variations increases as the background evaluation area becomes larger. Even in cases of extremely low background and low camera noise, the integration radius should not be chosen larger than about $3\sigma$ (1.5 times the lower limit), since the increased sensitivity to background variations outweighs any advantage from the marginally reduced grid-related noise.

Usually, the spot size is energy dependent. It increases towards lower energies because of less perfect focusing of the electron beam (phase space and space charge effects), but also due to the increasing influence of finite sizes of the domains or terraces on the sample surface. We therefore use an energy-dependent radius $r_\mathrm i$ of the integration disk
\begin{equation}
  r_\mathrm i^2 = r_\infty^2 + r_1^2/E \ , \label{eq:radius}
\end{equation}
where $r_\infty$ is the radius at very high energies (assumed to approach a constant value) and $r_1$ describes the increase of the radius towards low energies (for $E$ in electronvolts and $r_1 \gg r_\infty$, $r_1$ would be the radius at 1\,eV).  In the case of superstructure domains, the superstructure spots may be less sharp than substrate spots; this can be accounted for by choosing separate $r_1$ values for integer-order and superstructure spots. (In the ``Set Integration Radius'' input, for convenience, the user is asked to enter $r_\infty$ and the radius $r_\mathrm i$ at the lowest energy of the LEED image stack, not $r_1$.)
In ``azimuth blur'' mode, the semimajor axis of the integration (and background) ellipse is calculated by essentially the same equation, we only add $(\alpha_\mathrm{az} d_\mathrm{spot-(0,0)})^2$ to Eq.\ (\ref{eq:radius}). Here, $d_\mathrm{spot-(0,0)}$ is the distance between the spot of interest and the (0,0) spot, and $\alpha_\mathrm{az}$ is the blur angle in radians (assumed to be small, $\tan \alpha_\mathrm{az} \approx \alpha_\mathrm{az}$). Thus, the major axis of the integration ellipse is $\approx r_\mathrm i$ near the (0,0) spot and dominated by the $\alpha_\mathrm{az}$-dependent term for large distances from the (0,0) spot, as shown in Fig.\ \ref{fig:aperturePhotometry}(c).

\subsection{\label{sec:tracking}Spot tracking}

In a standard LEED experiment, the spots move radially with energy, with the distance from the (0,0) spot proportional to $1/\sqrt{E}$. Since spots can disappear over some energy range (if the intensity vanishes), it is necessary to take this motion into account when tracking the spots. Our approach starts at the energy where the fit for the spot pattern was obtained (Section \ref{sec:maxima}), and then continues searching at increasingly higher energies, thereafter descending the full energy range, and ascending again. At each energy, we make use of the spot pattern file to search for all spots in that file. Searching the full energy range both up and down makes sure that each spot will be tracked, provided that it can be found at any energy. When searching for spots that were not detected so far (or not detected at nearby energies), we follow two strategies, trying (i) the positions calculated by the polynomial fit in Eq.\ (\ref{eq:fitkxky}), scaled with $1/\sqrt{E}$, and (ii) a corrected position obtained from nearby spots already found. For these nearby spots, we calculate the deviations of their positions from the polynomial model. A linear fit of these deviations (as a function of $k_x$ and $k_y$, with weights decreasing with distance) yields a correction for the coordinates of the spot that we search for. This procedure corresponds to setting up a local coordinate system determined by the nearby spots, but still taking the overall nonlinear distortions of the LEED pattern into account.  The latter approach is especially valuable at very low energies, where the deviation between calculated and actual positions can be large (because of residual magnetic or electric fields, but also due to the large $1/\sqrt{E}$ scale factor between reciprocal-space coordinates and real-space positions). As soon as a given spot is found, its deviations from the polynomial model are kept for searching it at the next energies. If a spot has not been detected over a large energy range (default 30\,eV), these deviations may be unreliable and the polynomial model with corrections from the neighbors is also tried to find the spot (as if it were a spot never detected before). The code also includes plausibility checks for the spot positions. For example, the position is considered invalid if large jumps occur or if the position deviates too much from the polynomial fit. In case of doubt, uncertain positions are marked; these beams can be deleted from the analysis (``More$\gg$'' menu).

The spot positions obtained this way are smoothed in two passes. Smoothing uses a linear fit of the deviations from the polynomial distortion model in Eq.\ (\ref{eq:fitkxky}) as a function of $1/\sqrt{E}$, in a neighborhood of typically 30\,eV from each energy. Apart from the choice of $1/\sqrt{E}$ as the independent variable, this smoothing method is akin to a first-order Savitzky-Golay filter \cite{savitzky_smoothing_1964}.  To avoid a large impact of inaccurately determined spot positions near the intensity minima, the fit uses weights related to the spot significance, as introduced in Sec.\  \ref{sec:spotanalysis}. The first pass bridges gaps where a spot is invisible or only weak. (If there are no or not enough valid points at both sides, the energy range for fitting is extended to include enough points.) The linear fit also provides some extrapolation beyond the energy range where the spot was observed with sufficient significance to determine its positions. Extrapolation to high energies can sometimes lead to large slopes and, therefore, large deviations from the polynomial model. To avoid this problem it is beneficial to use an additional low-weight data point with zero deviation from the polynomial model for $1/\sqrt{E} = 0$. The second pass provides additional smoothing, with fit weights derived from the uncertainty of the fit results from the first pass. For obtaining the final (smoothed) spot positions, the fit results for the deviations are added to the positions calculated from the polynomial model in Eq.\ (\ref{eq:fitkxky}).

The choice of $1/\sqrt{E}$ as the independent variable in the linear fits is justified by our experience that the deviations from the polynomial model can be usually approximated as linear functions of $1/\sqrt{E}$. Thus, especially at low energies, this method provides much better results than smoothing methods not taking this $1/\sqrt{E}$-dependence into account.

The spot tracker also has a mode for LEED movies obtained at constant energy, typically used for analyzing a phase transition as a function of time or temperature. In this case, spot tracking may be necessary because thermal expansion of the sample or small movements of the sample holder due to its thermal expansion can cause the spots to move. This case is handled the same way as a standard tracking experiment, but without the $1/\sqrt{E}$ radial motion, and the image number replaces $1/\sqrt{E}$ as the independent variable in the smoothing of spot positions.

The spot tracker also features a LEEM (low-energy electron microscope) mode. In LEEM diffraction movies as a function of the energy, the spot pattern remains essentially stationary and spots move a few pixels at most. Thus, the LEEM mode works the same way as the analysis of LEED experiments at constant energy. Currently there are no provisions for automatic handling of the energy-dependent range of reciprocal space imaged by a LEEM instrument. (This would require a mask that changes with energy.) Therefore, if beams are invisible at low energies, their low-energy limit must be manually selected when editing the $I(V)$ curves.

\subsection{\label{sec:intensity}\textit{I}(\textit{V}) curve measurements}

The extraction of $I(V)$ curves uses aperture photometry (Section \ref{sec:spotanalysis}) at the smoothed spot positions described in the previous section. Having smoothly varying positions for the integration disk (and background) helps to obtain smooth $I(V)$ curves, without artifacts from jumps of the position of the integration area. As described in the previous section, these smoothed positions are also available for energy ranges with very low intensity of a given spot, where the images do not provide a reliable position.

Especially for superstructures, a frequent problem is having weak spots close to a strong one. In this case, the tails of the intensity distribution of the strong spot will lead to a curvature of the background intensity for nearby spots. Since we use a linear fit for the background, this can lead to apparently negative intensities of the weak spot. In this context, it is important that LEED diffraction maxima can be approximated by Gaussians only near the center (when ignoring the modulation by the grid). In the periphery we typically find a Lorentzian-like decay of the intensity with $1/r^2$, where $r$ is the distance from the center of the spot. This is in agreement with the expectation for kinematic scattering from phonons above the Debye temperature \cite{mckinney_thermal_1967}%
\footnote{The experimental results in Ref.\ \onlinecite{mckinney_thermal_1967} rather indicate an $1/r$ decay, which we cannot confirm.}.
The $1/r^2$ background implies that the tails of bright spots reach out rather far; this is the reason why they often affect the intensity measurement of nearby weak spots.
Therefore, our program provides an option to subtract the tails of the strong spots before measuring the weak ones. For this purpose, the spots are measured in order of decreasing intensity, and for each spot, after intensity measurement, a $1/r^2$ background is fitted in an annular region between $r_\mathrm i$ and $2r_\mathrm i$. This background is subtracted from the image in a large region around the spot before the next (weaker) spot is measured. In our experience, this background subtraction procedure eliminates the majority of minima reaching below zero in the $I(V)$ curves.

\subsection{\label{sec:asessment}Assessment of the data quality}

Besides the $I(V)$ curves, spot tracking produces a number of diagnostic plots, which help the user to assess the validity and quality of the data and optimize the choice of the parameters. One such plot shows the spot radii $\sigma$ (see Sec.\  \ref{sec:spotanalysis}) as a function of energy, together with a line at half the integration radius (separate for integer and superstructure, when different). This plot can be used to verify that the integration radius is chosen such that it is least $2\sigma$, as discussed in Sec.\  \ref{sec:spotanalysis}.

\begin{figure}
\includegraphics[width=8cm]{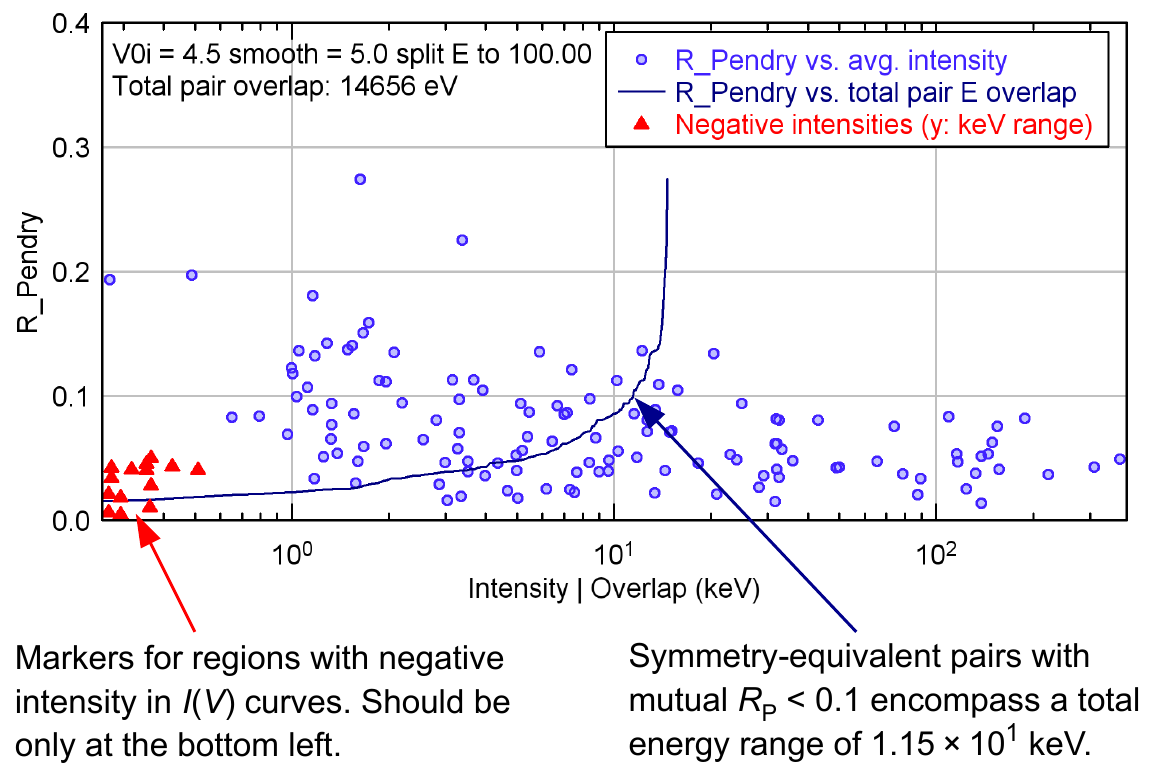}
\caption{\label{fig:spotTrackerPlot}Output plot of the spot tracker for assessing the quality of the $I(V)$ curves. This plot combines several aspects of the data. $R$-factors between pairs of symmetry-equivalent beams are blue dots. Typically, high-intensity beams (at the right) have lower $R$-factors than low-intensity ones. (The latter are more affected by the noise.) A summary of $I(V)$ curve regions with negative intensity is in red. For the red and blue data points, the $x$ axis is the intensity (average over $\approx 100$\,eV regions for the blue points of $R$ vs.\ intensity); the intensity is normalized such that the highest intensity in any $I(V)$ curve is 1000. The dark-blue curve ``$R_\mathrm{Pendry}$ vs.\ total pair $E$ overlap" allows a quick comparison of different data sets (lower is better). For this curve, the $x$ axis is the cumulative energy range of all pairs of symmetry-equivalent beams with an $R$ factor better than the $y$-axis value at that position of the curve. ({\sc ImageJ} currently does not support dual $x$ or $y$ axes, thus the double use of the axes.)} 
\end{figure}

A further plot (Fig.\ \ref{fig:spotTrackerPlot}) shows statistics useful to assess the quality of the $I(V)$ data
\footnote{Since the ``$I(V)$ Quality Statistics'' plot is based on the mutual $R$-factors of symmetry-equivalent beams, it is created only if there are at least two symmetry-equivalent beams. In case of non-normal incidence, for a valid result, only beams that are symmetry-equivalent at the given beam incidence should belong to the same group in the spot pattern file.}.
The blue points show Pendry's $R$ factor \cite{pendry_jphysc_1980} between pairs of symmetry-equivalent beams as a function of average beam intensity. (The $I(V)$ curves are smoothed for this using a 4\textsuperscript{th}-degree modified sinc smoother \cite{schmid_sg_2022}). Since beam intensities typically decrease with increasing energy, and the intensity affects the signal-to-noise ratio, the $I(V)$ curves are split into sections of $\approx 100$\,eV and each of these sections is analyzed and plotted individually.

The $I(V)$ data quality plot also includes an additional curve useful for judging the data quality (dark blue in Fig.\ \ref{fig:spotTrackerPlot}): This curve displays the total energy range of all pairs of symmetry-equivalent curves (split into $\approx 100$\,eV sections) where the $R$ factor does not exceed a given value (this value is the $y$ coordinate). The lower this curve, the better the agreement between equivalent beams. This curve is helpful for comparing data taken for the same system with different acquisition parameters or different spot tracking parameters. For instance, one can investigate the influence of the radius of the integration disk on the noise. Since the $x$ axis of this curve is the total energy range of the ``good'' data, most of this curve is not influenced by parameters that lead to elimination of the worst (e.g., most noisy) parts of the $I(V)$ curves. This makes it easy to obtain a valid comparison of data sets that include different energy ranges, such as one set containing low-intensity regions that are missing in the other set. (For comparison, curves can be copied from one {\sc ImageJ} plot to another by  ``Data$\gg$Add from plot...''.) 

The quality plot can also help when optimizing the voltage of the suppressor grid.  This can be done by comparing the mutual $R$-factors obtained from $I(V)$ movies acquired with different suppressor voltages. When the influence of electron capture and/or deflection by the grids on the $I(V)$ curves is minimized, the agreement between symmetry-equivalent beams is best. Since insufficient electron repulsion by the suppressor grid leads to an increase of the inelastic background, it is advisable to select the suppressor voltage on the strong-suppression side of the $R$ factor minimum, especially if the inelastic background is high and defocusing by the suppressor grid is not an issue.

Of course, comparing equivalent beams does not provide information on systematic effects that affect the equivalent beams the same way. For example, if the integration radius is too large, nonlinear variations of the inelastic background or the intensity tails of neighboring bright spots may affect the intensity of symmetry-equivalent curves in the same way; this cannot be detected via the $R$ factors between symmetry-equivalent curves. To diagnose at least one of these problems, the plot also contains statistics on $I(V)$ curve regions where the smoothed intensity is negative (red in Fig.\ \ref{fig:spotTrackerPlot}). The $x$ axis of these points gives the absolute value of the most negative intensity, and the $y$ axis gives the total energy range (per beam, in kiloelectronvolts) where the intensity is negative. Thus, high-quality data are characterized by few (or no) red points. If there are any red points, they should be close to the bottom left. If data obtained with an oval background (see Sec.\  \ref{sec:spotanalysis}) are badly plagued by negative intensities it is usually better to choose a circular background instead.

The plots generated when tracking spots also include  a set of selected $I(V)$ curves of symmetry-equivalent spots (useful to check alignment). There is also a plot stack (i.e., a set of plots) of the deviations of the spot positions from the polynomial model in Eq.\  (\ref{eq:fitkxky})  to check spot tracking. In addition, the user can plot a large number of quantities for a single beam, a few beams, or the overall measurement (available via the ``More$\gg$'' button of the spot tracker panel).

\subsection{\label{sec:I0}The beam current \textit{I}\textsubscript{0}}

Since the electron beam current $I_0$ usually changes during a LEED experiment, the raw intensities measured should be normalized by dividing by $I_0$. The beam current $I_0$ measured by LEED electronics can have an offset, which may be a substantial fraction of $I_0$ at low currents (especially for microchannel-plate LEED optics, where beam currents are a few orders of magnitudes lower than for standard LEED). This offset is named  $I_{00}$ and may depend on the energy; the program can subtract it from the measured $I_0$ values.

Any noise of $I_0$ will affect the final $I(V)$ curves. To avoid this problem, it is possible to smooth the $I_0 - I_{00}$ curve. These options are available via the ``Set Energies, I0, t'' button [Fig.\ \ref{fig:spottracker}(c); typical data are shown in Fig.\ \ref{fig:spottracker}(g)]. Smoothing uses a 4\textsuperscript{th}-degree modified-sinc kernel, which is similar to Savitzky--Golay filtering but provides better noise suppression and is less prone to overshoot at the boundaries \cite{schmid_sg_2022}.

In some cases, sudden jumps of the electron intensity occur.  (One possible reason is thermal expansion of some part of the electron source, leading to sudden movement when it overcomes static friction.) In such a case, smoothing of the electron current would smooth out the jump and result in an improper normalization. Such jumps can be also detected in the background intensity of the LEED images far from the spots, especially if the background is high (e.g., if the sample temperature is comparable to or higher than the Debye temperature). For these cases, the user can choose to take the rapid variations of the background (which tends to have low noise because it is the average over a large number to pixels
\footnote{As a background intensity, we use the average intensity of the 40\% darkest pixels in the screen area as defined by the mask, excluding the integration areas of the spots.})
and apply these variations to the smoothed $I_0$ values,
\begin{equation}
  I_0^\text{corr} = S(I_0 -I_{00}) \frac{I_\mathrm{b}}{S(I_\mathrm{b})}\ , \label{eq:i0ibcorr}
\end{equation}
where $I_\mathrm{b}$ is the background intensity  and $S$ represents the smoothing operator. 
The fraction ${I_\mathrm{b}}/{S(I_\mathrm{b})}$ in Eq.\ (\ref{eq:i0ibcorr}) is similar to a high-pass-filtered version of the background intensity with a baseline shifted to unity. If $I_\mathrm{b}$ is proportional to the  $I_0 - I_{00}$ (i.e., if $I_\mathrm{b}$ has the same energy dependence as the measured beam current), Eq.\ (\ref{eq:i0ibcorr}) ensures that the corrected $I_0^\text{corr}$ values will be proportional to $I_0 - I_{00}$; otherwise the slow variations of $I_0 - I_{00}$ are combined with the fast variations of $I_\mathrm{b}$.
In many cases, we find that the background intensity varies with energy in a manner similar to $I(V)$ curves, albeit with a lower relative amplitude of $\approx20$\%. A large portion of these background variations comes from stray light from very bright spots. If these variations are substantial, only rapid variations of the background intensity should be used for $I_0$ correction.  According to Eq.\ (\ref{eq:i0ibcorr}), this means that only mild smoothing should be used for $I_0 - I_{00}$ (e.g., a smoothing parameter of 10 points for 0.5\,eV energy steps
\footnote{The number of points entered as a smoothing parameter is the number of points of a moving-average filter with the same suppression of white noise. In other words, if  $n$ is entered as the number of points, the noise suppression factor is $1/\sqrt{n}$. The program calculates the filter kernel required for this noise suppression.}). 
The effect of the correction can be examined after spot tracking by plotting $I_0$ and $I_0^\text{corr}$ (available via the ``More$\gg$'' button).

\subsection{\label{sec:more}More spot-tracker features and utilities}

To enhance usability, the spot tracker contains additional features, available in the ``More$\gg$'' menu of the spot tracker panel. These include highlighting specific beams (to find and follow them easily in the LEED movie), and deleting beams in the output (fully or within some energy range). The ``More$\gg$'' menu also has an entry to list the current values of all parameters, together with the respective default values. The parameters are also written to a \texttt{.log} file upon saving data. The parameters can be also read from \texttt{.log} files, for processing the same or related data with identical parameters.

A further function of the spot tracker is undistorting one of the LEED images in the movie or  the full movie, based on the polynomial model in Eq.\ (\ref{eq:fitkxky}). The undistorted image stack can be created with a fixed $k$-space scale; then the spots do not move with energy. Averaging the images (``slices'') of such a stack over some energy range (or even the whole stack) provides an ``average'' LEED image with a high signal-to-noise ratio; also the adverse effects of the grids will be averaged out.

In addition, the package includes several utility plugins for handling $I(V)$ curves: averaging $I(V)$ curves, stitching curves with different (overlapping) energy ranges, resampling to a different energy step, (energy-dependent) intensity corrections, and $R$ factor calculations. A typical application of these utilities is mentioned in Sec.\ \ref{sec:preprocessing}: Noise reduction by averaging the $I(V)$ curves obtained with slightly different distance between the LEED optics and the sample. This is an efficient way to reduce the influence of the grids: In each movie, the electron beam of a given diffraction maximum reaches the grids at a slightly different position. Averaging uses the algorithm described in Sec.\  \ref{sec:curveeditor}, which includes smooth fading in or fading out if the energy ranges of the curves differ.

Most functions of the spot tracker and utilities can be controlled via the {\sc ImageJ} macro language. Automation is simplified by the {\sc ImageJ} Macro Recorder, which records the macro commands corresponding to a given workflow during manual operation.

\subsection{\label{sec:curveeditor}The \textit{I}(\textit{V}) curve editor}

\begin{figure*}
\includegraphics[width=16.5cm]{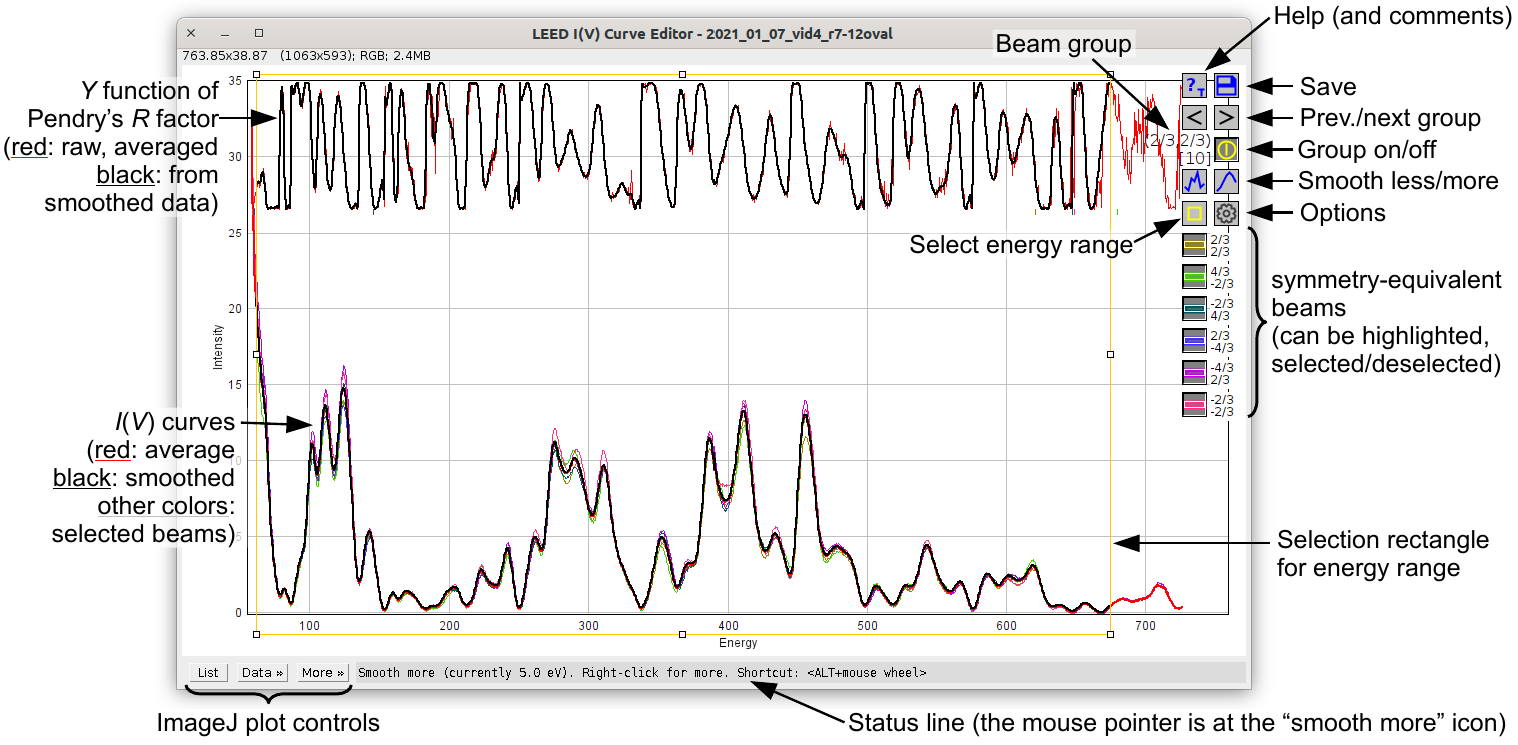}
\caption{\label{fig:curveeditor}Screenshot of the $I(V)$ curve editor. ``Group'' refers to a set of symmetry-equivalent beams. The buttons marked by arrows at the right side provide additional functionality with right-clicking (e.g., setting a parameter for all groups). Note that the intensity ($y$ axis) scale is chosen such that the highest spot intensity in the set of all $I(V)$ curves is $10^3$; even an intensity of 1 on this scale (0.1\% of the brightest spot) is sufficient for reasonable data quality. This $y$ axis scale only applies to the intensities, not to the $Y$ function of the $R$ factor, which is always shown in the same place at the top of the plot, irrespective of the y-axis scale.}
\end{figure*}

The $I(V)$ curve editor (Fig.\ \ref{fig:curveeditor}) is used for the final steps before the experimental $I(V)$ curves can be used for comparison with ``theoretical'' curves in structure optimization. These steps include averaging between symmetry-equivalent beams, selection of the useful data (sufficiently low noise, reasonable agreement between inequivalent beams), smoothing, and examination of the data. As described in the following, these steps are at least partly automated. This allows handling large data sets with hundreds of symmetry-inequivalent beams in a short time.

\emph{Curve averaging} ---
When averaging the $I(V)$ curves of symmetry-equivalent beams, the individual curves usually encompass different energy ranges. If a curve begins or ends within the energy range selected for the output, it is important to avoid jumps of the averaged intensity where that curve begins or ends. Therefore, before averaging, the curves are normalized and slow trends in the intensity ratio between the beginning and end of these curves are equalized. In addition, we use smooth fading in and/or fading out of the $I(V)$ curves that do not span the full energy range required for the output: Shorter curves use a linear increase or decrease of the weight in the averaging process at the respective end. (We use the imaginary part $V_\mathrm{0i}$ of the inner potential as an indication of a typical energy scale for variations in the $I(V)$ curves \cite{pendry_jphysc_1980}; e.g. the increase or decrease of the weight is over an energy interval of $4|V_\mathrm{0i}|$.)

\emph{Selection of data} ---
As mentioned in the introduction, a large database of experimental beams is important for a reliable structure analysis. [Therefore, the $I(V)$ curve editor displays the total energy range of symmetry-inequivalent beams selected in the status line at the bottom, unless the mouse pointer is at a button; then the status line displays information related to that button.] Selecting the data range is a compromise between a large database and rejecting low-quality data that increase the $R$ factor and do not help in the structure optimization.

As an aid for selecting the useful data, the $I(V)$ curve editor does not only plot  the original data and the (smoothed and unsmoothed) average over the symmetry-equivalent beams, but also the $Y$ function of Pendry's $R$ factor $R_\mathrm{P}$ \cite{pendry_jphysc_1980}. According to current knowledge, Pendry's $R$ factor is the method of choice for experiment--simulation comparison; in structure optimization it yields more accurate results than $R_2$, which is based on the squared difference of the normalized $I(V)$ curves \cite{sporn_accuracy_1998}. $R_\mathrm{P}$ is based on a comparison of the $Y$ function between experiment and theory. The $Y$ function is given by
\begin{equation}
   Y = \frac{L}{1+\left(V_\mathrm{0i}L\right)^2}\quad\text{with}\quad L=\frac{1}{I}\frac{\mathrm d I}{\mathrm d V}\ , \label{eq:pendryY}
\end{equation}
where $V_\mathrm{0i}$ is the imaginary part of the inner potential (typically, $|V_\mathrm{0i}| \approx 4$ to $5$\,eV) \cite{pendry_jphysc_1980}.
$Y$ is a nonlinear function of the logarithmic derivative $L$ of the $I(V)$ curves. Thus, it is not directly obvious to what degree the $Y$ function is influenced by noise and how it depends on smoothing. Plotting the $Y$ function allows the user to avoid data regions where $R_\mathrm{P}$ is strongly influenced by experimental noise and to examine the impact of smoothing on $Y$. 

Manual selection of the ``good'' beams and their useful energy ranges can be a cumbersome task, especially if there are hundreds of symmetry-inequivalent beams. The $I(V)$ curve editor therefore provides an option for automatic selection. The main basis for this analysis is an estimate of the noise of the $Y$ function, which is done by an $R$ factor-like comparison between the $Y$ function of the raw data and that after slight smoothing. As explained in Ref.\ \onlinecite{pendry_jphysc_1980}, the width of the features in an $I(V)$ curve is determined by $|V_\mathrm{0i}|$. A smoothing parameter of $0.55|\,V_\mathrm{0i}|$ leads to almost no noticeable change of low-noise $I(V)$ curves; therefore we use this smoothing parameter for the comparison with the $Y$ function of the raw curve. The comparison is made point by point and then smoothed by a running-average filter with a window length of  $2|V_\mathrm{0i}|$. With proper scaling, this procedure gives an estimate $r_\mathrm{n}(E)$ of the local contributions of the noise in the $I(V)$ data to the overall $R$ factor, assuming that the $R$ factor is dominated by noise
\footnote{Like the $R$ factor, the estimate of the local noise contribution $r_\mathrm{n}(E)$ is based on the squared difference of the $Y$ function of the unsmoothed and slightly smoothed $I(V)$ curve. To calculate the noise, we assume white Gaussian noise of the $Y$ function and make use of the known bandwidth of the smoothing filter \cite{schmid_sg_2022}. For the impact on the noise on the final $R$ factor we assume that a smoothing parameter of $1.0\,|V_\mathrm{0i}|$ will be chosen for the final curves.}.
[The user can select in the options menu to plot the noise function $r_\mathrm{n}(E)$.] Automatic selection of the beams and energy ranges (i) selects only curves or energy ranges where the average of the noise contributions $r_\mathrm{n}(E)$ is below a given limit $R_\mathrm{limit}$ and (ii) maximizes a figure of merit (FoM), given by
\begin{equation}
  F =\frac{E_\mathrm{max}-E_\mathrm{min}}{\langle r_\mathrm{n}(E)+c \rangle}
\end{equation}
where $E_\mathrm{min}$ and $E_\mathrm{max}$ are the bounds of the energy range selected for a given beam and the angle brackets denote the average over that energy range. 
For very low noise values $r_\mathrm{n}(E)$, the constant $c$ ensures that the FoM depends only on  the energy range, not on minor changes of the noise (we use $c = 0.5\,R_\mathrm{limit}$). Maximizing the FoM results in a large energy range, but penalizes energy regions with a high noise that would strongly increase the $R$ factor (high values of the denominator). For low-noise data, the denominator is dominated by the constant $c$, and only the energy range is maximized. A typical choice of the noise limit is $R_\mathrm{limit} \approx 0.05$; lower values are required for data with excellent agreement between calculated and experimental $I(V)$ curves ($R_\mathrm{P} \lesssim 0.1$) to avoid compromising the $R$ factor.

In addition to the noise-dependent part, automatic selection of beams and energy ranges also takes care of regions of negative intensity that result from uneven background. (We only consider negative intensities after smoothing, since noise may also cause negative intensities.) $R_\mathrm{P}$ is only defined for non-negative intensities $I$. To avoid negative values, a simple strategy is adding the absolute value of the most negative intensity to the $I(V)$ curve. Since $Y$ is a nonlinear function of the intensities (and their derivatives), such an upshift of the intensities affects $Y$ also in the regions where negative intensities do not occur. We therefore calculate the $R$ factor between the original and the upshifted curve in the positive-intensity regions. If this $R$ factor is too high (above $R_\mathrm{limit}$), instead of upshifting,  we exclude the negative-intensity region and its immediate vicinity.
The noise-dependent selection then proceeds as described above. Since we limit ourselves to $I(V)$ curves with a contiguous energy range, the FoM determines whether the energy range below or above the negative-intensity region is used. Typically, the ratio between background and intensities increases with energy, thus negative intensities occur more often at high energies and the low-energy side gets selected.

\emph{Noise-dependent smoothing} ---
For large data sets, noise-dependent adjustment of the smoothing parameter can be a lengthy task. By comparing the $R$ factor between experimental and simulated data, as well as by analyzing manually selected smoothing parameters, we found that a good choice is a smoothing parameter proportional to the average of the noise estimate $r_\mathrm{n}(E)$ of a given $I(V)$ curve raised to the power of 0.15
\footnote{Here, the smoothing parameter of the modified sinc filter \cite{schmid_sg_2022} is given in electronvolts and corresponds to the kernel length of a moving-average filter with equal suppression of white noise, as already described in Ref.\ \cite{Note11} above. This smoothing parameter is roughly proportional to the width of the smoothing kernel and inversely proportional to the bandwidth.}.

As a guide for choosing the smoothing parameter, the program also contains a facility to find the smoothing setting that optimizes the $R$ factor between smoothed experimental data and theoretical intensities (calculated, e.g., with {\scshape viperleed.calc} \cite{viperleed-calc}). When calculating such an optimal smoothing parameter for the whole data set or a subset thereof, the overall smoothing parameter can be adapted to the noise of each curve, as described above. The minimum of the $R$ factor against the smoothing parameter is rather shallow. To avoid oversmoothing, we do not use the value exactly at the minimum but the weakest smoothing that does not lead to an $R$ factor more than 1\% above the minimum.

\emph{Finding ``bad'' regions in the data} ---
For examination of the data quality, the $I(V)$ curve editor provides a function to selectively examine the cases of poor agreement of symmetry-equivalent beams with their average, based on Pendry's $R$ factor. It has to be noted that $R_\mathrm{P}$ is extremely sensitive to the exact shape of the curves at each minimum; even tiny deviations at the minima can lead to high values of the difference of the $Y$ functions, which determines the $R$ factor. Since such deviations are unavoidable, we first eliminate sharp peaks of the $Y$ function difference by a minimum filter \cite{burger_imageprocessing_2016}, followed by smoothing with a running-average filter. If the maxima of the local $R$ factor filtered this way exceed a user-defined threshold, they are flagged as regions of bad agreement. These regions are sometimes related to a beam passing over a defect of the LEED screen; in such a case the the affected beam can be excluded from averaging in this energy range.

\emph{Workflow} ---
The workflow for editing a large set of $I(V)$ curves can thus be reduced to (i) selection of a suitable noise limit $R_\mathrm{limit}$ and (ii) choice of the smoothing parameter for a curve with medium noise, followed by applying noise-dependent smoothing to all other curves. For some materials (notably 5d elements) or if very low energies (below 50\,eV) are included, the low-energy peaks may by substantially sharper than expected from the $V_\mathrm{0i}$ value. In such a case, weaker smoothing should be selected for beams including low energies. This can be done by first selecting the smoothing parameter for the first beams, which tolerate less smoothing due to their sharp peaks at lo energies. Then, starting with a ``higher'' beam (with the onset at higher energy), stronger smoothing can be applied to this beam and those further up. The final step in the workflow is the examination of the data quality; in many cases this can be restricted to the cases of poor agreement of symmetry-equivalent beams mentioned above (a small fraction of the whole data set).

When editing is complete, the final set of $I(V)$ curves can be saved into a file that is directly suitable as experimental input for structure optimization with {\scshape viperleed.calc} \cite{viperleed-calc}. The edit parameters (data ranges selected, smoothing strengths) are saved in an edit log file, which allows the user to (i) interrupt an editing session at any point and resume it later, (ii) modify a previous edit (changing the smoothing or adding/removing some beams), and (iii) apply the same editing parameters to a different data set with the same symmetry. This is useful, for instance, as a starting point when analyzing a later measurement of the same sample.

Apart from selecting and editing data, the $I(V)$ curve editor can be also used to compare different data sets, by opening two (or more) editor windows. Multiple editor windows are synchronized, which means that they show the same group of symmetry-equivalent beams over the same energy range (prior to any manual zooming). In addition, the $I(V)$ curve editor can be synchronized with the LEED movie shown in the spot tracker. Then, the energy selected in the spot tracker is marked in the $I(V)$ curve editor, and the beams selected in the $I(V)$ curve editor are marked in the spot tracker.


\section{\label{sec:conclusions}Conclusions}

We have developed an open-source software package
\footnote{The code is licensed under GNU GPLv3 or any later version. The documentation is licensed under Creative Commons Attribution CC BY 4.0.},
implemented as a set of {\sc ImageJ} plugins, for analysis of LEED movies, extraction and processing of $I(V)$ curves. The package was designed for (i) efficient and fast workflow and (ii) optimal data quality. For reaching these aims it includes many features not available in previous software solutions. The package currently contains about 15000 lines of code written over more than four years. The screenshots in Figs.\ \ref{fig:spotTrackerPlot} and \ref{fig:curveeditor} give an indication of the typical data quality obtained with our system. Even spots that have 0.1\% of the maximum spot intensity can be evaluated with acceptable noise (Fig.\ \ref{fig:curveeditor}). This data quality is substantially better than what could be achieved with legacy systems based on 8-bit images.

With our system, we have successfully performed spot tracking and $I(V)$ measurements for LEED movies acquired with four different experimental setups from more than 20 different structures: from simple Cu(111)-$(1\times 1)$ to complex ones with a dense arrangement of spots. The upper limit in complexity processed so far was a $(10 \times 10)$ superstructure on Pt(111) \cite{kisslinger_prb_2023}, where about 2000 $I(V)$ curves could be measured; about 350 inequivalent beams with sufficient quality were finally selected. For this structure, the proximity of spots limits the usable energy range to $E \le 400$\,eV. Taking advantage of parallelization for multi-core machines, the processing times for spot tracking and $I(V)$ measurements on a contemporary desktop computer are below one minute even for such a large data set. Including all user intervention (from opening the files and creation of the mask for a new LEED setup, setting all parameters, up to and including assessing the quality of the results), extraction of the raw $I(V)$ data from a typical LEED  movie takes about 10--15 minutes (less for subsequent similar movies with the same setup). The time required for selection, processing and examination of the data in the $I(V)$ curve editor is of the same order of magnitude. This has to be compared with a week of work for a structure with somewhat lower complexity \cite{von_witte_cdw_2019} when manually selecting and tracking spots one by one. Besides reduction of manual work (which also reduces the risk of human errors), our software contains many features that improve the data quality. Together with the other parts of the ViPErLEED project, we consider it an important step towards making LEED $I(V)$ studies more accessible.

\begin{acknowledgments}
The authors would like to thank Maximilian Buchta for providing test data for the ``azimuth blur'' mode and Alessandro Sala for providing LEEM image stacks for testing. This research was funded in part by the Austrian Science Fund (FWF) under doi 10.55776/F81, Taming Complexity in Materials Modeling (TACO). For the purpose of open access, the authors have applied a CC BY public copyright license to any Author Accepted Manuscript version arising from this submission. 
\end{acknowledgments}




\bibliography{viperleed-spottracker}

\end{document}